\documentclass[12pt]{article}
\usepackage{graphics}
\usepackage{latexsym}
\newcommand{\be}{\begin{equation}}
\newcommand{\ee}{\end{equation}}

\newcommand{\G}{\Gamma}

\newcommand{\ra}{\rightarrow}
\newcommand{\bea}{\begin{eqnarray}}
\newcommand{\eea}{\end{eqnarray}}

\newcommand{\we}{\wedge}
\newcommand{\sig}{\sigma}

\newcommand{\N}{{\cal{N}}}

\newcommand{\p}{\partial}
\def\id{\protect{{1 \kern-.28em {\rm l}}}}

\def\we{{\wedge}}

 \input amssym

\def\coeff#1#2{\relax{\textstyle {#1 \over #2}}\displaystyle}
\def\ZZ{\Bbb{Z}}

\def\IR{\Bbb{R}}
\begin{document}

\begin{titlepage}

\begin{flushright}
USC-04/03\\
UCLA/04/TEP/10\\
hep-th/0406145
\end{flushright}

\bigskip 
\bigskip
\centerline{\Large \bf A harmonic family of dielectric flow 
solutions   }
\medskip
\centerline{\Large \bf    with maximal supersymmetry}
\bigskip
\bigskip
\centerline{{\bf Iosif Bena$^1$ and Nicholas P. Warner $^2$}}
\medskip
\centerline{$^1$ Department of Physics and Astronomy}
\centerline{University of California}
\centerline{Los Angeles, CA  90095, USA }
\medskip
\centerline{$^2$ Department of Physics and Astronomy}
\centerline{University of Southern Califonia}
\centerline{Los Angeles, CA 90089, USA}
\bigskip
\centerline{{\rm iosif@physics.ucla.edu,~~ warner@usc.edu} }
\bigskip \bigskip

\begin{abstract}
We construct a new  harmonic family: dielectric flow 
solutions with maximal supersymmetry in eleven-dimensional supergravity. These 
solutions are asymptotically $AdS_4 \times S^7$, while in the infra-red 
the  M2 branes are dielectrically polarized into M5 branes. These solutions
are holographically dual to  vacua of the mass deformed  theory on M2 branes. 
They also provide an interesting insight on the supergravity solutions sourced 
by giant gravitons, allowing one to see how  supergravity solves
the giant graviton puzzle. 

\end{abstract}

\end{titlepage}

\section{Introduction}

Supersymmetry in the presence of branes is a well-studied subject,
and there are some rather general rules about how supersymmetry is
broken by combinations of different kinds of branes.   The simplest
of these rules involves harmonic distributions of intersecting branes in
which the supersymmetry is reduced by each set of branes through a
projection condition on the supersymmetry parameter.
The solution of  \cite{pw} was thus rather unexpected in that  it represents
an M-theory flow solution involving  both M2 and M5 branes and yet preserves
{\it sixteen}  supercharges.  This solution is asymptotic to  
$AdS_4 \times S^7$, and hence to a standard harmonic M2 brane solution
with sixteen supersymmetries, and yet it has non-trivial magnetic fluxes 
that dielectrically deform the M2 branes into M5 branes in the interior. It
does this while preserving all the supersymmetries, indeed it performs 
a dielectric rotation on the standard supersymmetry condition that defines
the supersymmetry for pure M2 branes.     

This new solution was obtained 
by ``lifting''  a four-dimensional gauged supergravity solution, and
while such methods have proven very powerful, they are limited in
two significant respects:  (i) Such lifts tend to be rather special examples
corresponding to higher levels of global symmetry, or very symmetric brane
distributions, (ii)  The lifting process does not yield much insight into
how to generalize the solutions and see if there are entire families 
of such solutions.   Our purpose here is to show that the solution in 
 \cite{pw} is part of a large family of new solutions, and that this family
 is characterized by a new form of ``harmonic'' distribution.  Indeed, it
 was not realized in  \cite{pw} that the new solution had an underlying harmonic
 structure. Moreover, given the recent work on algebraic Killing spinors
 \cite{spinor}, it is surprising that the new family of solutions is 
governed by a {\it linear} partial differential equation.
 
From the field theory perspective, we will see that the backgrounds 
we construct here are the duals of  vacua of the mass-deformed  theory on M2 branes. 
All these vacua have been  studied perturbatively in \cite{m2} and found to contain 
M2 branes polarized into  M5 branes. They are the eleven-dimensional 
analogues of the Polchinski-Strassler flow.

A first hint that there is indeed a whole family of solutions can be obtained by 
probing the  solution of \cite{pw}.   It turns out that the right probes are not 
M2 branes, as one might naively  expect, but M2 branes polarized into 
M5 branes \cite{m2}. By performing a 
probe analysis of this background, it is possible to see that there is a
configuration space  on which the potential for the probes is flat.  One should
thus be able to deform the particular solution of \cite{pw} by simply
adding more and more branes along this moduli space. 
It is interesting therefore to try to construct this family of
solutions directly. Although this might appear to be quite a daunting 
task, several methods have been recently developed \cite{spinor} that can be used 
to find this solution:  The basic idea is to find the  Killing spinors, 
and then use them to build the metric and forms.

We should remark that, {\it a priori}, it is quite amazing that such solutions 
even exist. Indeed, it seems to be commonly assumed that the only M-theory 
static\footnote{ Requiring the solution to be static excludes M-theory
giant gravitons and plane waves.} supergravity solutions with so much supersymmetry 
are Coulomb branch configurations of the  branes of M-theory or their duals. The solution 
of  \cite{pw}, and the
generalizations presented here are the first static solutions with maximal 
supersymmetry whose building blocks are not usual branes, but are
simply dielectric  branes \cite{myers}.    

Our presentation, to some degree, reverses the order in which the work was done:
We start in section 2 with the Ansatz  for the metric, the forms  and the Killing 
spinors.  To render the computation manageable we restrict to a family
of solutions that preserves an $SO(4)  \times  SO(4)$ subgroup of the
original $SO(8)$ ${\cal R}$-symmetry.  This means that we allow the
solutions to depend upon two ``radial'' variables, one for each of the two
$\IR^4$ factors in the $\IR^4 \times \IR^4 = \IR^8$ transverse to the 
original set of M2 branes.  The $SO(4)$'s then act as rotations on each of the
$\IR^4$'s.  This Ansatz enables us to find solutions with  arbitrary radial 
distributions of concentric shells  of branes in each  of these $\IR^4$'s.

The only other known dielectric brane configurations that preserve 
sixteen supercharges are  giant gravitons \cite{giant}. In fact, it is not hard 
to see that by reducing our  solutions along the M2 branes one obtains F1 strings 
polarized into D4 branes, which, when T-dualized along the strings, become 
momentum waves polarized into D3 branes in an 
external five-form field strength.  Hence, our solutions are dual to the giant
graviton 
solutions in type IIB, in the region near the $S^5$ equator, where the geometry has 
an $SO(4) \times  SO(4)$ ${\cal R}$-symmetry. It is quite likely that by starting 
from a more generic Ansatz one could use our methods to obtain the most 
generic giant graviton solution. 
However, our solutions describe giant gravitons for a rather large range 
of parameters, and are therefore very useful in trying to understand their 
physics. As we will see in section 5, our supergravity analysis indicates that 
valid giant graviton solutions only exists when the 
number of D3 branes is smaller than the square root of the number of gravitons, 
which strongly indicates that the solution to the giant graviton puzzle presented 
in \cite{douglas} is correct.

In section 3 we use some of the $G$-structure equations \cite{G-structure}
to simplify the Killing spinor Ansatz, and then in section 4 we complete
the solution. In section 5 we explore the physics of the solution when the M2 
branes are spread on a shell, and find that near the shell the solution 
reproduces the M5 branes solution. This 
links our solutions to giant gravitons, and to the perturbative analysis of \cite{m2}.
 Section 6 contains the brane-probe analysis of the
solution of \cite{pw}, and shows how this solution is related to
the linearized solutions of  \cite{m2}.  Section 7 contains 
some final remarks.

\section{The Ansatz}

Since we are imposing $SO(4)  \times  SO(4)$ on the 
$\IR^4 \times \IR^4$ transverse to the branes, we 
take the metric to have Lorentz frames of the form:
\bea
e^1 &=& e^{{A_0}(u,v)} dt ; \qquad e^2 = e^{{A_0}(u,v)} dx; 
~~~ e^3 = e^{{A_0}(u,v)} dy \\
e^4 &=& e^{{A_1}(u,v)} du ; \qquad  e^5 = e^{{A_2}(u,v)} dv \\
e^{5+j}  &=& \frac{1}{2} \, e^{{A_3}(u,v)}\,u \, \sig_j \qquad
e^{8+j}   =  \frac{1}{2} \, e^{{A_4}(u,v)}\,v  \,  \tau_j  \,, \\
\label{frames}
\eea
where $\sig_i$ and $\tau_i$ are the left-invariant 1-forms parameterizing 
the three-spheres invariant under the action of the SO(4) $\times$ SO(4) ${\cal
R}$-symmetry. 
Similarly, the most general 3-form potential  compatible with the ${\cal
R}$-symmetry is
\be 
C~=~ m_0(u,v) dt \wedge dx \wedge dy ~+~ m_1(u,v)\, \sigma_1 \wedge \sigma_2 
\wedge \sigma_3 ~+~ m_2(u,v) \, \tau_1 \wedge \tau_2  \wedge \tau_3 \,. 
\ee 

As in the solution \cite{pw}, the projector that determines the sixteen 
supersymmetries is a deformation of the usual M2 brane projector:
\bea
\Pi_0 &=&  {1 \over 2}(1 + p_1 \G^{123} + p_2 \G^{45678} + p_3 \G^{459\,10\,11} )\\
 &=& {1 \over 2}(1 + p_1 \G^{123} - p_2 \G^{1239\,10\,11} + p_3 \G^{1236\,7\,8} ) \,,
\label{Pi}
\eea
where we have used the fact that the product of all the gamma matrices is
the identity matrix.

Because only $C_{123}, C_{678} $, and $C_{9\, 10\, 11} $ are non-zero, it is not
hard to 
see that the combination 
\be
M \epsilon= \G^1 \delta \psi_1 +  \G^6 \delta \psi_6 + \G^9 \delta \psi_9
\label{m}
\ee
is independent of the Maxwell tensors. Indeed, if one drops 
all the derivative terms from the $\delta \psi_\mu$ in (\ref{m}), the 
remaining 16 $\times$ 16 matrix turns out to be 
\bea
M_0 &=& {1 \over 2}\frac{(1 + u\,{A_0}^{(1,0)} + 
    u\,{A_3}^{(1,0)} + 
    u\,{A_4}^{(1,0)})  }{\,
    e^{{A_1}}\,u} \G^4 \nonumber \\
&+& {1 \over 2} \frac{(1 + v\,{A_0}^{(0,1)} + 
    v\,{A_3}^{(0,1)} + 
    v\,{A_4}^{(0,1)})}{\,
    e^{{A_2}}\,v} \G^5 + \frac{e^{- A_3}}{2u} \G^{678}+ \frac{e^{- A_4}}{2v} \G^{9\,
10\, 11}
\eea

Up to now we have not used the reparametrization invariance available from the 
definitions of $u$ and $v$.  To fix one of these reparametrizations one can
essentially 
choose either $A_3$ or $A_4$, or a combination, at will.   To be more precise, one
can make a coordinate re-definition $\tilde u ={1\over 2L } u\, \exp(\Lambda + 
(A_0+A_3 +A_4))$,  $\tilde v ={1\over 2L }v\, \exp(-\Lambda +(A_0+A_3 +A_4))$, 
for some function $\Lambda(u,v)$, and a constant, $L$.  In the new metric, the
coefficient functions will satisfy $\tilde A_0+ \tilde A_3+\tilde A_4 = \log(4 L^2)$, 
but the change of variables will generate a $d\tilde u \, d \tilde v$ term.  
One can then choose the 
function $\Lambda(u,v)$ so as to eliminate this cross term.  One should note that
the pure M2 brane solution  has $A_3= A_4 =-{1 \over 2} A_0$, and in particular 
$A_0+ A_3+A_4$ is zero.  Thus the foregoing change of variables matches precisely
with the desired asymptotic behavior of our solution.  We have introduced a scale,
$L$, for
later convenience.  Dropping the tildes, we therefore fix the coordinate invariance, 
while preserving the form of the frames in (\ref{frames}), by setting:
\be
A_0+ A_3 + A_4 = \log(4 L^2) \,.
\label{A034}
\ee
 This simplifies $M_0$ to:
\be
M_0 = \frac{e^{{-A_1}}\,}{2\,
    u} \G^4 + 
\frac{e^{{-A_2}}\,}{2\,
    v}\G^5 + \frac{e^{- A_3}}{2u} \G^{678}+ \frac{e^{- A_4}}{2v} \G^{9\, 10\, 11}
\ee

Poincar\'e invariance means that the Killing spinors do not depend upon
the coordinates parallel to the brane.  Under the ${\cal R}$-symmetry 
the supersymmetries break up into various combinations of $SU(2)$ 
doublets.  This was analyzed in \cite{pw}, and one finds that 
the sixteen Killing spinors can be divided into two sets:   Half of the 
spinors are independent of all the sphere coordinates, while 
the other half depend in a very simple manner upon the sphere
coordinates.  To separate these two sets one needs to find a second
pair of projectors, $\Pi_1^\pm$,  that project onto these two sets of
eight Killing spinors.  Given these projectors, the  dependence of
the spinors upon the sphere directions is completely determined by the Lie
derivatives of the spinor, and one finds:
\bea
d\, \varepsilon &~+~&  \coeff{1}{2}\, 
\Big( \big(   \sigma_1\, \Gamma^{78} ~-~   \sigma_2\, \Gamma^{68}~+~
 \sigma_3\,  \Gamma^{67}\big) \,  \\
 & ~+~&  \big(   \tau_1\, \Gamma^{10\,11} ~-~  \tau_2\, 
\Gamma^{9\,11}~+~ 
\tau_3\, \Gamma^{9\,10}\big) \Big)\,\Pi_1^- 
 \, \varepsilon ~ ~=~ 0 \,, 
\eea
where $d$ is the exterior derivative on the spheres alone, $\Pi_1^-$ is the projector 
that annihilates the spinors that are independent of the sphere coordinates,
and is the identity matrix on the  second set of eight spinors.
It is elementary to find this projector $\Pi_1^-$:  Since the 
derivatives of the sphere-independent spinors do not enter 
the gravitino variations in (\ref{m}), $M_0$ must annihilate  these eight spinors. 
The  projector, $\Pi_1^-$, must therefore be the normalized multiple of $M_0$. Indeed,
one can take $\Pi_1^- = u\, e^{{A_1}}\,  \G_4 M_0$.  The condition that this
be a projector   ($\Pi_1^-  \,  \Pi_1^-  = \Pi_1^- $) means that one must have:
\be
\frac{e^{{-2 A_1}}\,}{4\,u^2} + \frac{e^{{- 2 A_2}}\,}{4\,v^2} = 
\frac{e^{- 2 A_3}}{4 u^2} + \frac{e^{- 2 A_4}}{4 v^2}
\label{link1}
\ee

In order not to further reduce the supersymmetry,  the projector
$\Pi_1^-$  should be compatible  ({\it i.e.} commute) with the projector (\ref{Pi}).  
This can only happen if
\be
{p_2 \over p_3} ={e^{A_4} v \over e^{A_3}  u}\,,
\ee
or
\be
p_2 = \frac{1}{2}\,\beta\,v\, e^{-{A_0} + {A_4}}~,\qquad
p_3 = \frac{1}{2}\,\beta\,u\, e^{-{A_0} + {A_3}}~\,, 
\label{proj}
\ee
for, {\it a priori}, some function, $\beta(u,v)$.  We will show below that $\beta$ is,
in fact, a constant.  Since $p_2$ and $p_3$ are the coefficients of
the non-trivial, magnetic components of the gauge field, the parameter,
$\beta$, represents the strength of the deformation away
from the pure M2 brane Coulomb branch.  Having fixed $p_2 $ and $p_3$,
 the remaining function, $p_1$,
 is obtained from requiring that $\Pi_0 \,  \Pi_0  = \Pi_0$, which implies:    
\be
p_1^2 ~+~ p_2^2  ~+~ p_3^2 ~=~1  \,.
\label{projnorm}
\ee

To proceed further it is useful to build
in much of what we have obtained so far by  introducing new variables 
$B_0,B_1,B_2,B_3$ so that  (\ref{A034}) is automatically satisfied:  
\bea
e^{A_0} &=& 4 L^2 e^{B_0}, \\
e^{A_1} &=& e^{B_1 - {B_0 \over 2}}, \\
e^{A_2} &=& e^{B_2 - {B_0 \over 2}}, \\
e^{A_3} &=& e^{B_3 - {B_0 \over 2}}, \\
e^{A_4} &=& e^{-B_3 - {B_0 \over 2}}, 
\eea

\section{The Killing spinors and the $G$-structure}

Having determined the projector, it is a straightforward exercise to 
obtain the eight Killing spinors that are annihilated both by $\Pi$ and by $\G_4
M_0$. 
The normalization of these spinors is most easily determined by using properties
of the $G$-structure.

First, if $\epsilon_i$ and $\epsilon_j$ are any two solutions to $\delta \psi_\mu =0$,
then \cite{G-structure} the following must be a Killing vector of the metric:   
\be
K_{ij}^{\mu} \equiv \bar\epsilon_i \G^{\mu} \epsilon_j \,.
\ee
We find that we can generate Killing vectors parallel to the branes, and
along the spheres in this manner.  The former fix the normalization of
the Killing spinors obtained from the projectors, while the latter Killing vectors 
are proportional to $\beta$, and the Killing equation implies that $\beta$
is constant.

Another very useful Killing spinor sandwich is  
\be
\Omega^{ij}_{\mu\nu} \equiv \bar\epsilon_i \G_{[ \mu } \G_{\nu ]}\epsilon_j 
\ee
which satisfies \cite{G-structure}:
\be
(d\Omega_{ij})_{\mu\nu\lambda} = K_{ij}^{\sigma}F_{\sigma \mu \nu \lambda} \,,
\ee
or in form language
\be
d \Omega = i_K F^{(4)}
\label{do}
\ee

If the components, $K^{\rho}$, are constant (as the Killing vectors we will use are), 
and the legs  of the form are not along sphere direction, one can integrate the
relation 
above  to obtain
\be
\Omega  = i_K C^{(3)}.
\label{co}
\ee

We can now chose a Killing spinor combination that gives a 
Killing vector whose only non-zero space-time component is  $K^3=-1$. 
This spinor combination also gives a two-form whose non-zero 
(space-time) components are:
\bea
\Omega_{12} &=& 64\,e^{3\,{B_0}}\,L^6\,  q_0 \\
\Omega_{34} &=& -2 e^{B_1 - B_2} L^2 u \beta \\
\Omega_{35} &=& 2 e^{B_2 - B_1} L^2 v \beta
\eea
where 
\be
q_0(u,v) \equiv {\sqrt{1 - \frac{e^{-3\,{B_0} + 2\,{B_3}}\,u^2\,{\beta }^2}{64\,L^4}
- 
      \frac{e^{-3\,{B_0} - 2\,{B_3}}\,v^2\,{\beta }^2}{64\,L^4}}}  \,.
\label{qzerodefn}
\ee
Combining this with (\ref{co}) determines 
\be
C_{123} = \Omega_{12} = 64\,e^{3\,{B_0}}\,L^6\, q_0 \,.
\ee 
In the limit $\beta \ra 0$ this reduces to the Coulomb branch solution. 

One can now use equation (\ref{do}) and the fact that $F^{(4)}_{45ij} =0$ 
to obtain a differential equation for the metric coefficients:
\be
\p_u \Omega_{35} = -\p_v \Omega_{34} \qquad  \Leftrightarrow  \qquad
\p_u\, (v\, e^{-(B_1 - B_2)}) =- \p_v\, (u\, e^{(B_1 - B_2)}) 
\label{bianchi1}
\ee

For the other components of the Maxwell tensor  things are a bit more complicated. 
In particular,  since they are along sphere directions, it is a little more
subtle to integrate (\ref{do}).  We can chose a combination of two Killing 
spinors that  gives rise to a Killing vector with
space-time components $K^{8} = K^{11}  = - \beta $, and we find:
\bea
\Omega &=&  e^{2\,{B_3}}\,L^2\,u^2\,  q_0 \, \sig_1 \wedge \sig_2 ~ -~  
e^{-2\,{B_3} }\,L^2\,v^2\, q_0 \,  \tau_1 \wedge \tau_2  ~+~  \nonumber  \\  
&& 2\, L^2 \Big(e^{2\,{B_3} }\,  u^2 + e^{-2\,{B_3} }\,  v^2\Big)^{-1}\, \Big[ 
 u\,\big(e^{{B_1} - {B_2} + 2\,{B_3}}\, u^2 + v^2 \, q_0) \, du \wedge \sig_3
\nonumber \\
 && ~-~
u^2\, v\,\big(e^{{B_1} - {B_2} + 2\,{B_3}} -  \, q_0) \, dv \wedge \sig_3 ~+~ 
 u\,v^2 \,\big(e^{{B_1} - {B_2} - 2\,{B_3}}\, - \, q_0) \, du \wedge \tau_3
\nonumber \\
 && ~-~
v\,\big(e^{{B_1} - {B_2} - 2\,{B_3}} \, v^2  -  u^2 \, q_0) \, dv \wedge \tau_3 
 \Big]  \label{Omdefn} 
\eea
 
Now we make some heavy use of (\ref{do}), and use the fact that:  
\be
d \sigma_1 = \sigma_2 \wedge \sigma_3\,, \qquad d \tau_1 = \tau_2 \wedge \tau_3\,, 
\qquad {\it and  \ cyclic.}
\ee
First we use the fact that the gauge field, $F$, has no components of the 
form $F_{45ij}$.  This means that all terms of the form $du\wedge dv \wedge 
\sig_3$ and $du\wedge dv \wedge  \tau_3$  must vanish in $d \Omega$, and 
the  result is two identities for partial derivatives of functions appearing in 
(\ref{Omdefn}).  These identities may be re-cast as integrability
conditions for two functions, $g_1(u,v)$ and $g_2(u,v)$, defined by:
\bea
\p_u g_1&=& 2\, \Big(e^{2\,{B_3} }\,  u^2 + e^{-2\,{B_3} }\,  v^2\Big)^{-1}\,
u\,\big(e^{{B_1} - {B_2} + 2\,{B_3}}\, u^2 + v^2 \, q_0) \label{gfnsdefna}   \\
\p_v g_1&=& - 2\,   \Big(e^{2\,{B_3} }\,  u^2 + e^{-2\,{B_3} }\,  v^2\Big)^{-1}\,
u^2\, v\,\big(e^{{B_1} - {B_2} + 2\,{B_3}} -  \, q_0) \label{gfnsdefnb} \\
\p_u g_2&=& 2\,   \Big(e^{2\,{B_3} }\,  u^2 + e^{-2\,{B_3} }\,  v^2\Big)^{-1}\,
u\,v^2 \,\big(e^{{B_1} - {B_2} - 2\,{B_3}}\, - \, q_0)  \label{gfnsdefnc} \\
\p_v g_2&=& - 2\,  \Big(e^{2\,{B_3} }\,  u^2 + e^{-2\,{B_3} }\,  v^2\Big)^{-1}\,
v\,\big(e^{{B_1} - {B_2} - 2\,{B_3}} \, v^2  +  u^2 \, q_0) 
\label{gfnsdefnd}
\eea
From (\ref{gfnsdefna})--(\ref{gfnsdefnd}) one can easily see that:
\be
\p_u (g_1 + g_2) ~=~ 2\, u \, e^{{B_1} - {B_2}  } \,, \qquad
\p_v (g_1 + g_2) ~=~- 2\, v \, e^{-({B_1} - {B_2})  } \,.
\label{easygs}
\ee
The integrability condition for this is precisely (\ref{bianchi1}), which is a nice 
consistency check.

With these definitions, one can then reconstruct $F$ and trivially integrate it to
obtain:
\bea
C^{(3)}  &=&  32\,e^{3\,{B_0}}\,L^6\, q_0  \, dt \wedge dx \wedge dy ~-~
{L^2 \over 2 \,\beta}\, \big(e^{2\,{B_3} } \, u^2 \,q_0 - g_1\big)\,
\sig_1\wedge\sig_2
\wedge\sig_3 \nonumber \\ 
&&~+~
{L^2 \over 2 \,\beta}\, \big(e^{- 2\,{B_3} }  \, v^2 \,q_0 + g_2\big)\,
\tau_1\wedge\tau_2
\wedge\tau_3 
\label{Cdefn}
\eea

At this point one has obtained as much as one can from the $G$-structure differential
equations for $K^\mu$ and $\Omega$, and so one resorts to solving the 
supersymmetry variations directly.  The problem is that the equations are still fairly
complicated, and so we make a simplifying assumption motivated by the solution of
\cite{pw}.  The {\it metric} of that solution was observed to have a discrete
symmetry, $\alpha \to
-\alpha$, $\theta \to {\pi \over 2} - \theta$, and the flux was odd under this 
symmetry provided that one also interchanged the $\sigma_j$ and $\tau_j$. 
In our formulation (see Section 2) this symmetry amounts to interchanging the 
$u$ and $v$ directions. The fact that the flux is odd under this symmetry 
suggests that the $u$ and $v$ directions must be undergoing
essentially identical, but conjugate processes.  Indeed, one can make this an exact
symmetry of the entire solution if one combines it with an orientation reversal
on the M2 branes and both $3$-spheres.  One of the consequences of this symmetry
is that one finds that $B_1(u,v) = B_2(u,v)$, and so we will now impose this condition
here.  The equations then simplify dramatically.

First, (\ref{bianchi1}) is trivially satisfied, and (\ref{easygs}) reduces to:
\be
g_1(u,v) = g(u,v) +  u^2 + k_1\,, \qquad
 g_2(u,v) = - (g(u,v) +  v^2 +  k_2)\,,
 \label{gonetwo}
\ee
for some function, $g(u,v)$, and where the $k_j$ are some integration constants.  
These constants are pure gauge in (\ref{Cdefn}), but they will be important in 
comparing our result to that of \cite{pw}.  Equation (\ref{link1}) becomes:
\be
 e^{{-2 B_1}} ~=~   (u^2 +v^2)^{-1}  \,\big (u^2 \,e^{ 2 B_3}  ~+~ 
 v^2 \,e^{ - 2 B_3} \big) \,, 
\label{nicerlink1}
\ee
while equations (\ref{gfnsdefna})--(\ref{gfnsdefnd})  now reduce to:
\bea
\p_u g &=&-  2\, (u^2 +v^2)^{-1}  \,u \, v^2\,  e^{{2 B_1}} \,
\big(e^{ -2\,{B_3}}\, - \, q_0) \label{gdefna}   \\
\p_v g &=& - 2\, (u^2 +v^2)^{-1}  \,  u^2\, v\, 
e^{{2 B_1}} \, \big(e^{ 2\,{B_3}} -  \, q_0) \label{gdefnb}  
\eea

The strategy is now very simple:  One first uses  (\ref{nicerlink1}) 
to eliminate all appearances of $B_1$ and its 
derivatives from the supersymmetry variations.  One then 
solves (\ref{nicerlink1}), (\ref{gdefna}) and (\ref{gdefnb}) 
for $B_0$ and $B_3$ in terms of $\p_u g$ and $\p_v g$ (remember
that $q_0$ contains $B_0$), 
and then uses this to  eliminate all appearances of $B_0$,
$B_3$, and derivatives, from the supersymmetry variations.
As a consequence, every supersymmetry variation can be written
in terms of the first and second derivatives of $g$.

It is then trivial to find the condition on $g$ that solves the entire system,
and the  rather amazing result is that one has sixteen supersymmetries
if and only if $g$ satisfies the {\it linear} equation:
\be
{\p^2 g \over \p u^2} +{\p^2 g \over \p v^2} -{1 \over u}{\p g \over \p u} 
-{1 \over v}{\p g \over \p u} = 0
\label{har1}
\ee
Even more surprisingly,  this equation implies that the function 
$h \equiv u^{-2} v^{-2} g $ satisfies
\be
{\p^2 h \over \p u^2} +{\p^2 h \over \p v^2} +{3 \over u}{\p h \over \p u} 
+{3 \over v}{\p h \over \p u} ~=~ {1 \over u^3}\, \p_u \big(u^3 \, \p_u h)
~+~ {1 \over v^3}\, \p_v \big(v^3 \, \p_v h) = 0,
\label{har}
\ee
which is the harmonic equation in $\IR^8$! 

We have also verified that this solution of the supersymmetry variations
solves all the eleven-dimensional equations of motion.

\section{The solution}

We have found that any function, $h(u,v)$, satisfying (\ref{har}) 
gives a solution to the eleven-dimensional supergravity equations of motion, 
preserving sixteen supercharges. However, it is not {\it a priori} clear that 
the harmonic function $h(u,v)$ 
becomes the usual Coulomb branch harmonic function when $\beta \ra 0$. 
We will first find the relation between the non-dielectric harmonic function 
and the one we have here, and then explore several examples that make the physics 
of the solutions transparent. 

The first example is  the harmonic function that has a point source. 
When $\beta \ra 0$, this becomes the usual $AdS_4 \times S^7$ solution.
A non-zero $\beta$ corresponds to a perturbation of the 
$AdS_4 \times S^7$ solution by a non-normalizable mode. In the dual boundary 
theory, this perturbation is dual to turning on an operator that gives 
equal masses to the four chiral multiplets (8 fermions and 8 bosons). In fact 
this perturbation is none other than the perturbation studied in \cite{m2}, 
which causes the M2 branes to polarize into M5 branes. 

We can also spread the source of the harmonic function on one of the two 3-spheres 
in the problem.  This gives a solution that  looks like an M5 brane when
one is near the 3-sphere. In the UV this solution asymptotes to the pure M2
brane solution.   Therefore, this  solution is the exact solution sourced by M2 branes 
polarized into M5 branes, which 
is dual to one of the vacua of the massive boundary theory.  Since the equation
satisfied by the master function $g$ is linear, the 
exact supergravity solutions dual to all the vacua of this theory 
(containing multiple shells of M2 branes polarized 
into M5 branes) can be trivially found by superposing the harmonic functions of 
individual shells.

\subsection{The general dielectric solution}

The maximally supersymmetric harmonic solutions we found can be expressed in terms 
of the function $g$ obeying (\ref{har1}), using equations 
(\ref{qzerodefn},\ref{Cdefn},\ref{nicerlink1},\ref{gdefna},\ref{gdefnb}):
\bea
e^{4\,{B_3}} &=&
  \frac{v^2\,\left( 2\,L^2\,u^2 + 
       v\,g^{(0,1)} - 
       u\,g^{(1,0)} \right) }{u^2\,
     \left( 2\,L^2\,v^2 - 
       v\,g^{(0,1)} + 
       u\,g^{(1,0)} \right) } \\
e^{4\,{B_1}} = e^{4\,{B_2}}&=& 
  \frac{\left( 2\,L^2\,u^2 + 
       v\,g^{(0,1)} - 
       u\,g^{(1,0)} \right) \,
     \left( 2\,L^2\,v^2 - 
       v\,g^{(0,1)} + 
       u\,g^{(1,0)} \right) }{4\,
     L^4\,u^2\,v^2} \\
e^{-3\,{B_0}}  &=& \frac{L^2}{\beta^2}
  \frac{2\,L^2\,v\,g^{(0,1)} + 
     2\,L^2\,u\,g^{(1,0)} - 
     (g^{(0,1)})^2 -
     (g^{(1,0)})^2}{2\,u\,v\,
     {\sqrt{2\,L^2\,u^2 + 
         v\,g^{(0,1)} - 
         u\,g^{(1,0)}}}\,
     {\sqrt{2\,L^2\,v^2 - 
         v\,g^{(0,1)} + 
         u\,g^{(1,0)}}}} \\
{1 \over 2}C_{123} &=& 
  \frac{64\,L^4 \beta^2 \,u\,v\,
     \left( -2\,L^2\,u\,v + 
       u\,g^{(0,1)} + 
       v\,g^{(1,0)} \right) }{-2\,L^2\,v\,g^{(0,1)} - 
     2\,L^2\,u\,g^{(1,0)} + 
     (g^{(0,1)})^2 + 
     (g^{(1,0)})^2} \\ 
\beta C_{678} &=& g - 
  \frac{L^2\,u\,\left( u^2 + v^2 \right)
       \,g^{(1,0)}}{2\,L^2\,v^2 - 
     v\,g^{(0,1)} + u\,g^{(1,0)}} \\
- 
\beta C_{9\, 10\, 11} &=& g - 
  \frac{L^2\,v\,\left( u^2 + v^2 \right)
       \,g^{(0,1)}}{2\,L^2\,u^2 + 
     v\,g^{(0,1)} - u\,g^{(1,0)}}
\eea

We observe that in order to recover the usual harmonic solution in the limit $\beta
\ra 0$, 
the function $g$ should go to zero like $\beta^2$. Therefore, when $\beta = 0$, the 
harmonic function which gives the Coulomb branch solution is:
\be
Z = {1 \over 128 L^4 \beta^2}\, {u {\p_u g } + v {\p_v g} \over u^2 v^2}
\label{z}
\ee
One can easily show that the operator $ u {\p_u g } + v {\p_v g}$ maps harmonic
functions to harmonic functions, and so, if $g$ satisfies (\ref{har1}), the function
$Z$ is 
harmonic in $\IR^8$. Therefore, if one wants to find the solution that is asymptotic
to a 
Coulomb branch solution given by $Z(u,v)$, one needs to invert (\ref{z}) to find $g$.

For this, it is useful to introduce the new coordinates $x, y$ defined by
\be
u=e^{x+y},~~~~v = e^{x-y}
\label{xy}
\ee
Equation (\ref{z}) becomes:
\be
 128 L^4 \beta^2 Z(x,y) e^{4x} =  { {\p g(x,y) \over \p x}}
\ee
and can be integrated straightforwardly to yield the results we quote below
(\ref{g1},\ref{g2}).

In fact, the condition (\ref{z}) only determines $g(u,v)$ up to an arbitrary 
additive function $g_1(u,v)$ satisfying (\ref{har1}) and  $u {\p_u g_1 } + v {\p_v
g_1} =0$.
It is not hard to see that up to an irrelevant additive constant, 
the only function satisfying these two equations is 
\be
g_1(u,v)= c\, {u^2 -v^2 \over u^2 + v^2} \,.
\ee
Hence, for any value of the constant $c$ the function $g+g_1$ gives a 
solution to the equations of motion. We should note that all components 
of the solution depend explicitly on $g_1$. Thus, given any Coulomb 
branch solution there exists a one parameter family of dielectric solutions 
which asymptote to that solution in the limit $\beta \ra 0$.

\subsection{ Coincident branes: the $AdS_4 \times S^7$ flow with no polarized branes}

When all the branes are located at a single point, the harmonic function $Z$ is 
\be
Z = {R^6 \over r^6} = {R^6 \over {(u^2 + v^2 )^3}},
\ee
where $R$ is the $AdS$ radius. Equation (\ref{z}) implies 
\be
u \p_u g + v \p_v g = 128 L^4 \beta^2 R^6  {u^2 v^2 \over {(u^2 + v^2 )^3}},
\ee
and hence $g$ is given by
\be
g =  {B u^2 v^2 \over {(u^2 + v^2 )^3}} + g_1
\label{g1}
\ee
where $B \equiv - 64 L^4 \beta^2 R^6  $

If we set $g_1$ to zero, the metric coefficients and the forms become:
\bea
e^{-3\,{B_0}} &=& \frac{L^2\,B\,
     \left( L^2\,{\left( u^2 + v^2 \right) }^4 + B\,\left( u^4 - u^2\,v^2 + 
v^4 \right)  \right) }
     {{\left( u^2 + v^2 \right) }^3\,
     {\sqrt{3\,B\,v^2\,\left( u^2 - v^2 \right)  +L^2\,
          {\left( u^2 + v^2 \right)}^4}}\,{\sqrt{L^2\,
          {\left( u^2 + v^2 \right)}^4 - 3\,B\,\left( u^4 - u^2\,v^2
            \right) }}\,{\beta }^2} \nonumber \\
e^{4\,{B_3}} &=&  \frac{3\,B\,v^2\,
      \left( u^2 - v^2 \right)  + 
     L^2\,{\left( u^2 + v^2 \right)
          }^4}{L^2\,
      {\left( u^2 + v^2 \right) }^4 -
      3\,B\,\left( u^4 - 
        u^2\,v^2 \right) } \nonumber \\
e^{4\,{B_1}} &=&\  \frac{\left( 3\,B\,v^2\,
        \left( u^2 - v^2 \right)  + 
       L^2\,{\left( u^2 + v^2 \right)
            }^4 \right) \,
     \left( L^2\,
        {\left( u^2 + v^2 \right) }^4
        - 3\,B\,
        \left( u^4 - u^2\,v^2 \right)
           \right) }{L^4\,
     {\left( u^2 + v^2 \right) }^8} \nonumber \\
{1 \over 2}  C_{123} &=&  \frac{-32\,L^4\,
     {\left( u^2 + v^2 \right) }^3\,
     \left( L^2\,
        {\left( u^2 + v^2 \right) }^4 - B\,\left( u^4 - 4\,u^2\,v^2 + v^4 \right) 
       \right) \,{\beta }^2}{B\,\left( L^2\,{\left( u^2 + v^2 \right) }^4
        + B\,\left( u^4 - u^2\,v^2 + 
          v^4 \right)  \right) } \\
{1 \over 2} C_{678} &=&  \frac{-\left( B\,u^4\,
       \left( 3\,B\,v^2\,
          \left( u^2 - v^2 \right)  -
          2\,L^2\,
          {\left( u^2 + v^2 \right)
            }^4 \right)  \right) }
     {2\,{\left( u^2 + v^2 \right) }^
      3\,\left( L^2\,
        {\left( u^2 + v^2 \right) }^4
        - 3\,B\,
        \left( u^4 - u^2\,v^2 \right)
           \right) \,\beta } \nonumber \\
{1 \over 2} C_{9\, 10\, 11} &=&  \frac{B\,v^4\,
     \left( -2\,L^2\,  {\left( u^2 + v^2 \right) }^4- 3\,B\,
        \left( u^4 - u^2\,v^2 \right) 
    \right) }{2\,{\left( u^2 + v^2 \right) }^3\,\left( 3\,B\,v^2\,
    \left( u^2 - v^2 \right)  + L^2\,{\left( u^2 + v^2 \right) }^4 \right) 
     \,\beta }\nonumber 
\eea

It is not hard to see that this solution is quite sick for very small $u$ and $v$. 
Some of the metric coefficients become negative, and the size of the $\sigma$ and 
$\tau$ circles also becomes zero, which signals very large curvature. We should 
also note that one cannot remove this singularity by adding  $g_1$ or
modifying the coordinates. The function $g_1$ is much smaller than 
$g$ in the region where the solution becomes pathological.  Moreover,
the only harmonic functions with a source at $u=v=0$ are proportional to 
${1 \over r^6}$, and so   the only way to soften the singularities
involve smearing out the branes.   We will consider this below.
 
In the ultraviolet, the leading asymptotic terms of the  foregoing expressions 
can be easily found, remembering that $B \propto \beta^2$ : 
\bea
e^{-3 \,{B_0}} &\ra& \frac{L^2\,B}
   {{\left( u^2 + v^2 \right) }^3\,
     {\beta }^2} = 64 L^6 Z \\
e^{4\,{B_3}} &\ra& 1 \\  
C_{123} &\ra& \frac{-64\,L^4\,
     {\left( u^2 + v^2 \right) }^3\,
     {\beta }^2}{B} = - {Z^{-1}} \\
C_{678} &\ra&  -  \frac{2 B\,u^4}
     {{\left( u^2 + v^2 \right) }^3\,
       \beta }  \label{c1} \\
C_{9\, 10\, 11} &\ra&  - \frac{2 B\,v^4}
     {{\left( u^2 + v^2 \right) }^3\,
       \beta },
\label{c2}
\eea
which is an $AdS_4 \times S^7$ solution perturbed with transverse fluxes, 
whose  magnitude is
\be
C_{678} \sim u^4 \beta Z.
\ee

These fluxes are exactly the ones found in the analysis of the gravity dual of the 
mass deformed theory on a large number of coincident M2 branes \cite{m2}. 
They correspond to $AdS_4$ non-normalizable modes, and are dual to turning 
on a mass for the four chiral multiplets of this theory. 
This is therefore the M-theory version of Polchinski-Strassler
\cite{Polchinski:2000uf}. 
As discovered in the perturbative analysis in \cite{m2}, the transverse flux causes 
the M2 branes to  polarize into M5 branes, exactly as in the IIB case 
\cite{Polchinski:2000uf}. Moreover, in \cite{Polchinski:2000uf} all the vacua 
of the gauge theory were matched to brane configurations, and it was argued 
that the solution with no polarized branes should not be physical, because there
is no corresponding  gauge theory vacuum. Our analysis confirms that this intuition 
also extends to the M-theory case. We will come back to the issue of this 
solution when we discuss giant graviton physics in section 5.

Since the perturbative analysis of the system reveals the possibility of having 
M2 branes  polarized into M5 branes, it is quite natural to expect the solutions to
be in 
the class of solutions obtained here. This is further supported by the 
probe analysis performed in section 6.

\section{M2 branes polarized into M5 branes, and giant graviton physics}

In order to find the function $g(u,v)$ that gives the solution corresponding to 
all the M2 branes polarized into one M5 brane shell, we use the fact that if we 
take $\beta \ra 0$, we expect this solution to 
become the Coulomb branch solution given by M2 branes smeared on a 3-sphere. 
It is therefore natural to expect that at least in a region of space, the function
$g_0$ 
that gives the Coulomb branch solution to give the full polarized solution. 
It is however also possible that the desired function,  $g$, also receives corrections
at higher order in $\beta$:   $g = g_0 + \beta^2 g_2 + \beta^4 g_4 + ... $. 

We start by finding and analyzing the solution to lowest order, and neglecting
higher order corrections in $\beta$. 
The harmonic function $Z$ that gives a Coulomb branch distribution of 
M2 branes on a 3-sphere of radius $R_0$ is given by
\be
Z = \frac{ R^6 }
  {{\left( R_0^2 - 2\,R_0\,u + u^2 + 
        v^2 \right) }^{\frac{3}{2}}\,
    {\left( R_0^2 + 2\,R_0\,u + u^2 + 
        v^2 \right) }^{\frac{3}{2}}} ~~, 
\ee
where $R$ is the $AdS$ radius. Using the change of variables (\ref{xy}) 
and inverting (\ref{z}), one finds $g$: 
\be
g = D  {u^2 - R_0^2 - v^2 \over {{\left( R_0^2 - 2\,R_0\,u + u^2 + 
        v^2 \right) }^{\frac{1}{2}}\,
    {\left( R_0^2 + 2\,R_0\,u + u^2 + 
        v^2 \right) }^{\frac{1}{2}}    }  } + g_1,
\label{g2}
\ee
where $ D = {16 L^4 R^6 \beta^2 \over R_0^2 } $. It is not hard to check that 
$g$ satisfies (\ref{har1}). We would like to explore whether the 
solution near the shell has M5 brane form. The contribution of $g_1$ near the shell 
is small, so we only explore the solution for $g_1=0$. To see the form of the
solution near 
the shell we substitute (\ref{g2}) in the solution, and define $x \equiv u-R$. 
In the near shell region one has  $x \ll R$. 

We first focus on the combination $2 L^2 u^2 - u g^{(1,0)} + v g^{(1,0)}$,
which appears in the numerator of $e^{4 B_3}$. Near the shell, this is approximately:
\be
2 L^2 u^2 - u g^{(1,0)} + v g^{(1,0)} \approx 2 L^2 R_0^2 - {D R_0 v^2 \over (x^2 +
v^2)^{3/2}}
\label{sing}
\ee 
For $x,v \sim {D \over R_0 L^2}$ this combination becomes negative. This may 
signify either a coordinate breakdown, or the fact that the naive function 
$g$ receives corrections at higher order in $\beta$.  To see the M5 brane 
behavior it is better to stay away from this region, but remain near the shell. 
We therefore, work in a region in which 
\be
{D \over L^2 R_0} \ll x,v \ll R_0.
\label{range}
\ee
In this region the second term of the right hand side of 
(\ref{sing}) can be ignored.

In addition to (\ref{sing}),  our solution involves another potentially singular
expression 
in terms of the harmonic function.   One can estimate the combination in the numerator 
of $e^{-3 B_0}$ to see that:
\be
2\,L^2\,v\,g^{(0,1)} + 2\,L^2\,u\,g^{(1,0)} - (g^{(0,1)})^2 -
(g^{(1,0)})^2 \approx {2 L^2 D v^2 R_0 \over (v^2+x^2)^{3/2}}- {D^2 v^2 \over
(v^2+x^2)^2} 
\label{sing2}
\ee
This combination also becomes negative for $x,v \sim {D \over R_0 L^2}$, but the
singular locus,  where it becomes negative is different from that of  (\ref{sing}). 
Hence, the singularity is not merely a coordinate artifact, but probably signifies 
the fact that the harmonic function which gives the dielectric brane distribution 
is not straightforward to obtain starting from the harmonic function which gives 
the Coulomb branch distribution:   There may indeed be higher order corrections
in $\beta$.

Fortunately, in the regime (\ref{range}) the second term of (\ref{sing2}) 
drops out, and all the metric functions are positive. The solution becomes:
\bea
e^{4 B_1} &=& {D R_0 \over 2 L^2 (x^2+v^2)^{3/2}} \nonumber \\
e^{4 B_3} &=& \left({D R_0 \over 2 L^2 (x^2+v^2)^{3/2}}\right)^{-1}  \label{5met} \\
e^{-3 B_0} &=& {D^{1/2}  L^3 \over \sqrt{2} R_0^{3/2} \beta^2  (x^2+v^2)^{3/4}   }
\nonumber
\eea
In order for the solution near the shell to match the M5 brane 
harmonic solution, we need the metric along the branes  
(in the $1,2,3,\sigma_1, \sigma_2,\sigma_3$ directions) to be 
proportional to $Z_5^{- 1/3}$, and the metric transverse to the 
M5 branes (in the $u,v, \tau_1, \tau_2, \tau_3$ directions) to be 
proportional to $Z_5^{2/3}$, where $Z_5 \sim  {Q_5 \over (x^2+v^2)^{3/2}} $. 
This gives four equations for the three functions $B_0,B_1,$ and $B_3$, 
which have the solution: 
\be  
 e^{4 B_1} = Z_5~, ~~~~  e^{-4 B_3} = Z_5~, ~~~~ e^{-6 B_0} = Z_5
\ee
Hence, after a trivial rescaling of the coordinates, (\ref{5met}) gives the $AdS_7 \times S^4$ 
metric sourced by a collection of M5 branes at $u=R_0$. The radius of the 4-sphere is  
\be
R^2_{S^4} = v^2 e^{-2 B_3-B_0}|_{x=0} = \left({D \over 2 \beta} \right)^{2 \over 3}~,
\ee
giving the number of M5 branes in the shell
\be
N_5 = {8 L^4 R^6 \beta \over \pi l_p^3 R_0^2} 
\label{Q5}
\ee

One can also compute this charge by integrating the $\tau$ field strenghts on a 4-sphere 
around the shell. Their leading behavior in the near-shell region is:
\bea
F_{u\tau_1\tau_2\tau_3} &=& - {3 D \over 2 \beta} {v^4 \over (v^2+x^2)^{5/2}} \\
F_{v\tau_1\tau_2\tau_3} &=& {3 D \over 2 \beta}  {v^3 x \over (v^2+x^2)^{5/2}} 
\eea
We introduce spherical coordinates around the shell $x= \rho \cos \theta, ~
v = \rho \sin \theta$ and integrate $F^{(4)}$ on the 4-sphere, to find:
\be
(2 \pi l_p)^3 N_5 = \int_{S^4} F^{(4)} = {2 D \over \beta} \times V_{\Omega_3}  = {64 \pi^2
L^4 R^6 \beta \over R_0^2} ,
\ee
in perfect agreement with (\ref{Q5}). 

Hence both the metric and the forms show the presence of an M5 brane 
wrapped on a three-sphere at radius $R_0$. Our solution is therefore the 
first supergravity solution that exactly describes a situation in which branes are
polarized into other branes wrapped on a 3-sphere. 

For a given M5 dipole charge $N_5$ the radius of the polarization shell is
\be
R_0^2 = {16 \pi l_p N_2 \beta \over  N_5}
\ee
where $N_2$ is the total number of M2 branes.
This formula agrees beautifully with the perturbative analysis in 
\cite{m2}, where it was found that in the external fields (\ref{c1},\ref{c2}) 
the radius of the M5 brane shell is also
\be
R_0^2 \sim \beta {N_2 \over N_5}.
\ee

Thus, as we have expected, our solutions describe the backgrounds dual 
to the vacua of the mass deformed M2 brane worldvolume theory. To obtain 
the dual of the most general such vacuum, corresponding to M2 branes 
polarized into several concentric M5 shells in both transverse $\IR^4$'s, one simply needs
to superpose the harmonic functions sourced by the shells.

It is interesting to examine what happens in the region where 
(\ref{sing}) and (\ref{sing2}) become negative. The fact that these combinations 
vanish on different loci, most probably indicates the fact that the function 
$g$ in (\ref{g2}) gets corrected near the shell.  
Even if our solution is not valid for $x,v \sim {D \over R_0 L^2}$, one can argue 
very strongly that in this region it should have a simple continuation to 
the precise M5 brane form. Indeed, for a very large region outside 
$x,v \sim {D \over R_0 L^2}$ the metric and the forms  give exactly the 
$AdS_7 \times S^4$ solution of supergravity. Moreover, 
we know from the supergravity solutions of dissolved branes 
\cite{Costa:1996zd} that very close to the M5 
branes the effect of the  dissolved M2 branes becomes very small, and the 
metric is exactly the  M5 brane metric. Hence one should be able to find
a set of harmonic corrections, perhaps combined with a simple
coordinate re-definition that continues the 
$AdS_7 \times S^4$  to the region with $x,v \sim {D \over R_0 L^2}$.

The foregoing argument that there should be a smooth transition to
the pure M5 brane solution was contingent upon there being a large 
intermediate region in which the solution did indeed become that of the M5 brane.
Such a region only exists if  (\ref{range}) is satisfied, and this 
requires that $R_0^2 \gg D$.   In terms of the M5 brane and 
M2 brane charges, Eq. (\ref{range}) is equivalent to:
\be
N_2 > N_5^2
\label{cond}
\ee
which is the same as the validity condition of the Polchinksi-Strassler perturbative 
analysis!  When (\ref{cond}) is not satisfied, the interior of the solution cannot
be interpreted 
in terms of an M5 brane shell.

\subsection{How supergravity solves the giant graviton puzzle}

Two of the observations we made in the previous subsection have an important role in 
the physics of giant gravitons. The first is that a solution for M2 branes polarized 
into an M5 brane only exists when the number of M2 brane is greater 
than the square of the number of M5 branes. The second is that the 
solution in which the hamonic function is sourced by coincident M2 branes is 
not physical.

As we have discussed, M2 branes polarized into M5 branes can be 
related to $AdS_5 \times S^5$ giant gravitons in a region near the 
$S^5$ equator.  This
is seen through a  dimensional  reduction along the M2 brane direction 
and a T-duality along the resulting F1 string.  As it is well known, giant gravitons
in 
$AdS_5 \times S^5$ correspond to field theory chiral primaries.  As discussed in 
\cite{cjr,vijay} using the $\N=4$ Super Yang Mills, sphere giants and $AdS$ giants 
correspond to maximally anti-symmetric and respectively maximally symmetric 
chiral primaries.   This was expanded upon in \cite{douglas} (by using the auxiliary 
theory describing the gravitons in their near horizon region), where 
it was argued that a field theory chiral primary is dual to {\em two} 
giant gravitons, but these two dual gravitons have non-overlapping regimes of 
validity. More precisely,  a field theory state corresponding to $J/k$ sphere 
giant gravitons of angular momentum $k$ each is also dual to $k$ $AdS$ giant 
gravitons of angular momentum $J/k$ each. 
 
A crucial ingredient in this argument was the fact that these two giant 
gravitons cannot exist simultaneously. In the  equatorial region (which is the 
region where our solutions are valid), the $AdS$ and sphere giants are related by a 
$\ZZ_2$ symmetry interchanging the two $SO(4)$'s. The fact that the gravitons 
have non-overlapping ranges of validity translates in the requirement that a 
giant graviton  exists only if the number of D3 branes and the angular 
momentum satisfy
\be
N_3 <  \sqrt{J}.
\ee
Since the number of D3 branes becomes the M5 brane charge, and $J $ becomes 
the M2 charge, this condition is precisely the one in equation (\ref{cond}).

Another important aspect of solving the giant graviton puzzle is finding the 
fate of the solution with unpolarized gravitons. This is nothing but the singular
solution 
discussed in section 4.  Of course, we have not argued that {\em all } solutions 
without giant gravitons are singular. However, the most obvious families of
solutions 
which contain coincident branes are singular. The most likely way to make them 
non-singular is to smear the harmonic functions; however this precisely 
corresponds to introducing giant gravitons. Other work on finding supergravity duals of
giant graviton systems has appeard in \cite{silva}.

\section{M5 branes with M2 charge as probes for small $\zeta$}

In this section we show that the solution obtained in \cite{pw} is a part of our 
family of solutions. We also probe this solution using M2 branes polarized 
into M5 branes, and show that the probes have a moduli space. This was, in fact,
our first indication that  the solution in \cite{pw} is  a member of a 
family of solutions, and this was the original motivation for this work.

\subsection{Obtaining the gauged supergravity solution}

It is elementary to recover the solution of \cite{pw}.   The deformation parameter,
$\beta$, is related to the corresponding parameter, $\zeta$, in  \cite{pw}  via
\be
\beta ~\equiv~ {1 \over L} \, \sin \zeta\,.
\ee
The relevant change of variables is 
\be  
u~=~ {e^{-{1\over 2} \,\alpha}   \over \sqrt{ \sinh \alpha}} \, \cos \theta \,, \qquad
 v~=~{e^{{1\over 2} \,\alpha}   \over \sqrt{ \sinh \alpha }}  \, \sin \theta  \,. 
\ee 
Note that this is precisely the change of coordinates used in \cite{pw}
to convert the harmonic flow with $\beta =0$ to the manifestly
harmonic form in terms of $u$ and $v$.

If one substitutes the metric functions of  \cite{pw} into  (\ref{gdefna})
and (\ref{gdefnb}), and converts them into differential equations with respect
to $\alpha$ and $\theta$, then they can be trivially integrated to obtain:
\be
g~=~\sin^2\Big(\frac{1}{2}\,\zeta\Big) \,  \cos 2\theta \,,
\ee
It is thus this very simple  ``master function'' that determines the 
entire solution in  \cite{pw}.  To recover the solution exactly as presented
in this reference one must take $k_1 = \sin^2({1 \over 2} \zeta)$ and
$k_2 = - \sin^2 (\frac{1}{2} \zeta)$ in (\ref{gonetwo}) so as to make 
$g_1 \sim \cos^2 \theta$ and $g_2 \sim \sin^2 \theta$.  As we underlined
before, the choice of these constants is pure gauge.

Thus, we see that the solution of \cite{pw} is a single example
of the very large class of solutions obtained here.

\subsection{The action of the M2-M5 probes}

Because of the complexity of the general solution, we will only perform
the analysis  when $\zeta$ is small.   This will be sufficient to exhibit
the polarization effects and the moduli space.   
In this limit the background \cite{pw} consists of a distribution of parallel M2 branes, 
with a very small M5 dipole moment. It is thus natural to use a spherical test M5
brane with a large M2 charge to probe this background.  

In order to give an M5 brane an M2 charge one needs to turn on its
world-volume self-dual three-form field strength. The M5 action is rather
complicated \cite{Perry:1996mk,Aganagic:1997zq,Pasti:1997gx}, but it is not  
intractable.   The main complication arises from the self-duality of the 3-form, which 
implies that this form will always have one leg
along the time direction.  Hence the Lagrangian and the Hamiltonian will
be different. Moreover, self duality makes finding the Hamiltonian rather messy. 

Instead of pursuing this route, we use the fact that compactified
eleven-dimensional supergravity is type IIA supergravity in ten dimensions. 
The probe potential is independent of the compactification radius. 
Thus, it is given by the potential of a spherical D4 brane with large F1 charge,
 in the background \cite{pw} reduced along $e^2$.

One can give F1 string charge to a D4 brane by turning on its world-volume
field strength $F_{01}$. Suppose that the D4 brane is wrapped on a
3-sphere whose vielbeins are $e_{\psi_1},e_{\psi_2}$ and $e_{\psi_3}$. The
Born-Infeld Lagrangian is simply:
\be
L = - e_{\psi_1}e_{\psi_2}e_{\psi_3} e^{-\Phi}\sqrt{(e_0 e_1)^2 -(B_{01} +
F_{01})^2}.
\ee
The quantized F1 string charge in a non-trivial background is:
\be
\Pi = {\delta L \over \delta F_{01}} = e_{\psi_1}e_{\psi_2}e_{\psi_3}
e^{-\Phi} {B_{01} + F_{01} \over \sqrt{(e_0 e_1)^2 -(B_{01} + F_{01})^2}} =  N
\ee
This gives 
\be
{1\over \sqrt{(e_0 e_1)^2 -(B_{01} + F_{01})^2}} = {1 \over e_0 e_1} \sqrt{ 1+ 
{N^2 e^{2 \Phi}\over e^2_{\psi_1}e^2_{\psi_2}e^2_{\psi_3}  }}
\ee
The Hamiltonian is 
\bea
H &=& \Pi F_{01} - L = e_{\psi_1}e_{\psi_2}e_{\psi_3} e^{-\Phi} { e^2_0 e^2_1 - 
B_{01}(B_{01} + F_{01})\over \sqrt{(e_0 e_1)^2 -(B_{01} + F_{01})^2} } \\
& = & e_0 e_1 \sqrt{  e^2_{\psi_1}e^2_{\psi_2}e^2_{\psi_3} e^{-2 \Phi} + 
N^2 } - N B_{01}  \,, 
\label{N}
\eea
and we may think of this as the effective potential of the spherical D4 brane
with an F1 charge equal to $N$.

A few consistency checks: In the background created by a large number of F1 
strings, the leading
terms in the large  $N$ expansion cancel -- this is simply a reflection of the fact that
parallel fundamental  strings do not interact. 
For $N=0$ we recover the D4 brane action. For $B=0$ the masses of the 
D4 brane and F1  string add in quadratures. Moreover one can obtain this 
expression for the Hamiltonian of a D-brane with fundamental string charge by using 
S-duality. The action of a D3 brane with D1 brane 
charge $N$ in a RR background field $C^{(2)}$ is $ e_0 e_1\sqrt{e_2^2 e_3^2 e^{-2 \Phi}+ 
N^2 e^{-2 \Phi}}- N C_{01} $ 
and after S-duality becomes exactly the IIB version of (\ref{N}).

Remembering that the eleven-dimensional vielbeins are related to the 
ten-dimensional ones by:
\be
e|_{10} = e^{\Phi/3}e|_{11}, ~~~~ e_2|_{11}= e^{2 \Phi /3} 
\ee
and using (\ref{N}) it is not hard to reexpress the effective potential as a
function of the eleven-dimensional parameters:
\be
H_{11} =  e_1 e_2 e_3 \sqrt{  e^2_{\psi_1}e^2_{\psi_2}e^2_{\psi_3} + N^2 } - N
C_{012} ,
\label{h11}
\ee
where now the vielbeins are eleven-dimensional.

The Hamiltonian of the D4 brane also contains a Wess-Zumino coupling to the
background RR-fields and NS-NS fields:
\be
H^{\rm D4}_{WZ} = - C_{5} + B_2 \wedge C_3.
\ee
To lift this to eleven dimensions it is convenient to act on it with an exterior derivative, 
and use the IIA equation of motion, $d C_5 = - * F_4 + H_3 ~\wedge C_3$, to obtain:
\be
d H^{\rm D4}_{WZ} = * F_4 + B_2 \wedge F_4
\ee
When reducing the background \cite{pw} to ten dimensions, the $B$ field descends 
from $C_{012}$, while $F_4$ and $* F_4$  descend from the four-form $F^t$ transverse 
to the M2 branes. Hence, the WZ coupling 
of a probe M5 brane in the background \cite{pw} is:
\be
d H^{M5}_{WZ} = * F_4^t + C_3^{(123)} \wedge F_4^{t}
\label{HM5}
\ee

\subsection{The probe action}

In the limit of large string charge and  small $\zeta$, the square roots in
(\ref{h11})
can be expanded, giving three contributions to the effective potential:
\bea
H_1 &=& {1 \over 2N} {e_1 e_2 e_3  e^2_{\psi_1}e^2_{\psi_2}e^2_{\psi_3}
}\\
H_{2} &=& H^{M5}_{WZ} \\
H_3 &=& N(e_1 e_2 e_3 - C_{123}   ) = N e_1 e_2 e_3- {N k^3 X_3 \over 2 \sinh^3
(\alpha)} 
\eea
Both $H_1$ and $H_3$ can be found rather straightforwardly. Suppose that
the directions of the 3-sphere are be given by $e^{\psi_i} = \cos \gamma_i
e^{\sigma_i} + \sin \gamma_i e^{\tau_i}$.  The angles, $\gamma_i$ thus
represent the orientation of the probe along the two geometric 3-spheres.
For the moment we will take all the $\gamma_i$ to be equal: 
 $\gamma_i = \gamma$.

For small $\zeta$, 
\bea
H_1 &= & {k^3 L^6 \over 4 N} {(\cos \theta \cos \gamma e^{- \alpha/2} +
\sin \theta \sin \gamma e^{ \alpha/2})^6  \over \sinh^3 \alpha }\\
H_3 &=& {N k^3 \zeta^2 \over 4} {\cos^2 \theta e^{-\alpha} + \sin^2 \theta
e^{\alpha} \over \sinh \alpha } 
\label{Hthree}
\eea

Nevertheless, to find $H_2$ we need to do a bit of work. We Hodge dualize the
forms in \cite{pw}, compute $d H_{WZ}$ using (\ref{HM5}), and integrate it to obtain: 
\bea
H_2 &=& {-k^3 L^3 \zeta \over 2 } {\cos^4 \theta  e^{-2 \alpha} \over \sinh^2 \alpha} 
~dx^1 \we ~dx^2 \we ~dx^3 \wedge \sigma_1 \we \sigma_2 \we \sigma_3  \nonumber \\  
&+ & {-k^3 L^3 \zeta \over 2 } {\sin^4 \theta  e^{2 \alpha} \over \sinh^2 \alpha} 
 ~dx^1 \we ~dx^2 \we ~dx^3 \we \tau_1 \we \tau_2 \we \tau_3 
\eea

To be supersymmetric, the effective potential must come from a 
superpotential, and hence must have the form
of a sum of perfect squares of analytic functions in the
fundamental physical parameters.   One can see that  
$H=H_1+H_2+H_3$  has such a form 
only if $\theta=0$ or $\theta=\pi/2$. For these two angles the effective
potential can be conveniently written:
\bea
H_{\rm total} &=&  {k^3 L^6 \over 4 N } {\cos^2 \theta e^{- \alpha} \over 
\sinh \alpha}\left(  \cos^2 \theta \cos^3 \gamma e^{- \alpha} - \zeta N /L^3 
\right)^2 \\
 &+&{k^3 L^6 \over 4 N } {\sin^2 \theta e^{ \alpha} \over \sinh \alpha}
 \left( \sin^2 \theta \sin^3  \gamma e^{ \alpha} - \zeta N/L^3  \right)^2
\label{h}
\eea
For general  $\gamma_i$'s, $\cos^3 \gamma$ is replaced by 
$\cos \gamma_1 \cos \gamma_2\cos 
\gamma_3$ and similarly for $\sin^3 \gamma$.

This suggests  that the only supersymmetric ways to add a brane
probe is along the $u$ or $v$ axis ($\theta =0$ or $\theta =\pi/2$).
Moreover, it shows that there is a vanishing effective potential, and thus a non-trivial
moduli space, when the orientation angles, $\gamma_i$,
of the probe appropriately adjust as the radial coordinate,
$\alpha$ varies.  This moduli space
is therefore six-dimensional:  The three $\gamma_j$ with one radius
and a constraint yield half of the moduli space, and the
other half is given by the rotations of the brane around the
3-sphere that it wraps.

We end this section by observing that for $\zeta \ne 0$ there is no moduli
space for a probe M2 brane unless it carries M5 charge. If the probe
has no M5 brane charge then $H_1$ and $H_2$ vanish, and the effective 
potential is simply given by  $H_3$ (\ref{Hthree}).  This is
manifestly positive, and hence there is no moduli space. This shows that the 
building blocks of the solution \cite{pw} are not usual M2 branes, but M2 branes 
polarized into M5 branes. Hence, the probe computation illustrates
one of the conclusions of   section 5, by revealing the dielectric structure 
of these solutions.

\section{Conclusions and Future Directions}

We have used some of the rather powerful Killing spinor methods that have been 
developed over the past few years \cite{spinor,G-structure} to construct a harmonic family
of maximally supersymmetric eleven-dimensional supergravity solutions 
describing M2 branes dielectrically  polarized into M5 branes. 

All other dielectric supergravity solutions that have been constructed until now
\cite{other} contain branes polarized into branes one dimension higher, wrapped on a 
circle. Polarizations of codimension two or three, which are physically much more 
interesting, have proven difficult to obtain via usual solution generating techniques. 
Our solution is the first of this kind. 

The physics of our solution is very relevant for illuminating a few rather puzzling 
phenomena which have appeared in systems containing polarized branes. One of these 
systems is the giant graviton. Our solutions are dual to giant gravitons near the 
$S^5$ equator, and show that only giant gravitons with 
angular momentum greater than the square of the number of D3 branes are physical. 
Moreover, our solutions support the fact that Kaluza Klein gravitons 
with large angular momentum are unphysical. This provides further evidence
for the  fact that in any parameter regime there exists only one supergravity 
solution dual to a  field theory chiral primary.

Our solutions are the M-theory analogue of the yet to be found type IIB 
solutions dual to the $N=1^*$ theory in four dimensions \cite{Polchinski:2000uf}. 
Many vacua of this theory undergo a confinement--deconfinement phase transition. 
In supergravity this corresponds to a transition from a configuration of 
D3 branes polarized into a shell of  D5 branes, to D3 branes polarized into a 
shell of NS5 branes. However, the 
transition regime is inaccessible to Born-Infeld analysis used in 
\cite{Polchinski:2000uf}. To describe this regime, one really needs the full 
supergravity solution. In the M-theory solution constructed here this
deconfinement phase transition corresponds to going from an M5 brane shell of 
charge $k$ on the $\sigma$ sphere to an M5 brane of charge $N/k$ on the $\tau$ sphere 
as $k$ becomes larger than $\sqrt{N}$. Our class of solutions should capture all 
intermediate solutions, and thus can be used to understand how this 
transition takes place. It would be very interesting to find the function $g$ 
giving the phase transition supergravity solutions, and to use these solutions 
to learn more about this transition in gauge theory.

More generally, we believe the methods used in this study may well prove
useful in finding the exact type IIB duals of the $N=1^*$ theory. Since these solutions 
also describe concentric shells of polarized branes \cite{Polchinski:2000uf}, 
it is quite likely that they will also be generated by a  
master function satisfying a linear equation. 

Last, but not least, our solutions put to rest the well-established belief that 
the only maximally supersymmetric static solutions are Coulomb branch distributions of 
branes. The fact that supergravity allows maximally supersymmetric solutions built out 
of polarized branes attests to the richness of its physics, and to its fascinating 
interaction with string theory.

\subsection*{Acknowledgments}

This work was supported in part by funds provided by the DOE under 
grants  DE-FG03-84ER-40168 and DE-FG03-91ER-40662, and by the NSF under 
grants PHY00-99590 and PHY01-40151.  \footnote{ Any opinions, findings and 
conclusions expressed in this material
are those of the authors and do not necessarily reflect the views
of the U.S. Department of Energy or of the National Science Foundation.}

\end{document}